\documentclass{scienceadvhack}

\usepackage{scicite}
\usepackage{times}
\usepackage{graphicx}
\usepackage{amsmath}
\usepackage{amssymb}

\newcommand{\kepler}{{\it Kepler}}

\newcounter{lastnote}

\title{Magnetically gated accretion in an accreting ``non-magnetic'' white dwarf}

\author
{Simone Scaringi,$^{1\ast}$
Thomas J. Maccarone,$^{2}$
Caroline D'Angelo,$^{3}$
Christian Knigge,$^{4}$
Paul J. Groot.$^{5}$ 
\\
\\
\normalsize{$^{1}$School of Physical and Chemical Sciences, University of Canterbury, Christchurch 8041, New Zealand}\\
\normalsize{$^{2}$Department of Physics and Astronomy, Texas Tech University, Box 41051, Lubbock, TX 79409–1051, USA.}\\
\normalsize{$^{3}$Leiden Observatory, Leiden University, Leiden 2300RA, The Netherlands}\\
\normalsize{$^{4}$School of Physics and Astronomy, University of Southampton, Highfield, Southampton SO17 1BJ, UK.}\\
\normalsize{$^{5}$Department of Astrophysics/IMAPP, Radboud University, PO Box 9010, 6500 GL Nijmegen, The Netherlands}\\
\\
\normalsize{$^\ast$Corresponding author. E-mail: simone.scaringi@canterbury.ac.nz}
}

\date{}

\begin{document}

\maketitle 

\begin{abstract}
White dwarfs are often found in binary systems with orbital periods ranging from tens of minutes to hours in which they can accrete gas from their companion stars. In about 15\% of these binaries, the magnetic field of the white dwarf is strong enough ($\geq 10^6$ Gauss) to channel the accreted matter along field lines onto the magnetic poles$^{1,2}$. The remaining systems are referred to as ``non-magnetic'', since to date there has been no evidence that they have a dynamically significant magnetic field. Here we report an analysis of archival optical observations of the ``non-magnetic'' accreting white dwarf in the binary system MV Lyrae (hereafter MV Lyr), whose lightcurve displayed quasi-periodic bursts of $\approx 30$ minutes duration every $\approx 2$ hours. The observations indicate the presence of an unstable magnetically-regulated accretion mode, revealing the existence of magnetically gated accretion$^{3-5}$, where disk material builds up around the magnetospheric boundary (at the co-rotation radius) and then accretes onto the white dwarf, producing bursts powered by the release of gravitational potential energy. We infer a surface magnetic field strength for the white dwarf in MV Lyr between $2 \times 10^4 \leq  B \leq 10^5$ Gauss, too low to be detectable by other current methods. Our discovery provides a new way of studying the strength and evolution of magnetic fields in accreting white dwarfs and extends the connections between accretion onto white dwarfs, young stellar objects and neutron stars, for which similar magnetically gated accretion cysles have been identified$^{6-9}$. 
\end{abstract}

MV Lyr spends most of its time in an optically bright ($m_V \approx 12$) luminosity state. Occasionally and sporadically (typically about once every few years) it drops by more than a factor 250 in brightness for short durations (weeks to months), sometimes fading to $m_V \approx 18$ (Fig 1a). Other accreting white dwarfs show similar optical brightness variations, and fall under the class of so-called nova-like variables$^{10-13}$. The physical mechanism for these sudden drops in brightness is not well established$^{14-16}$. As the luminosity of these systems is dominated by the release of gravitational potential energy of the gas in the disk, it is clear that the brightness variations are a direct consequence of changes in the mass-transfer rate through the accretion disk in these systems: during the bright phases (``high states''), the mass-transfer rate can be as high as $\gtrsim 10^{-8} M_{\odot}/\text{yr}$, whilst during the faint phases (``low states''), the mass transfer rate can drop as low as $\lesssim 10^{-11} M_{\odot}/\text{yr}$ (refs $12,13$).

\begin{figure*}[h]
\begin{center}
\includegraphics[width = \textwidth]{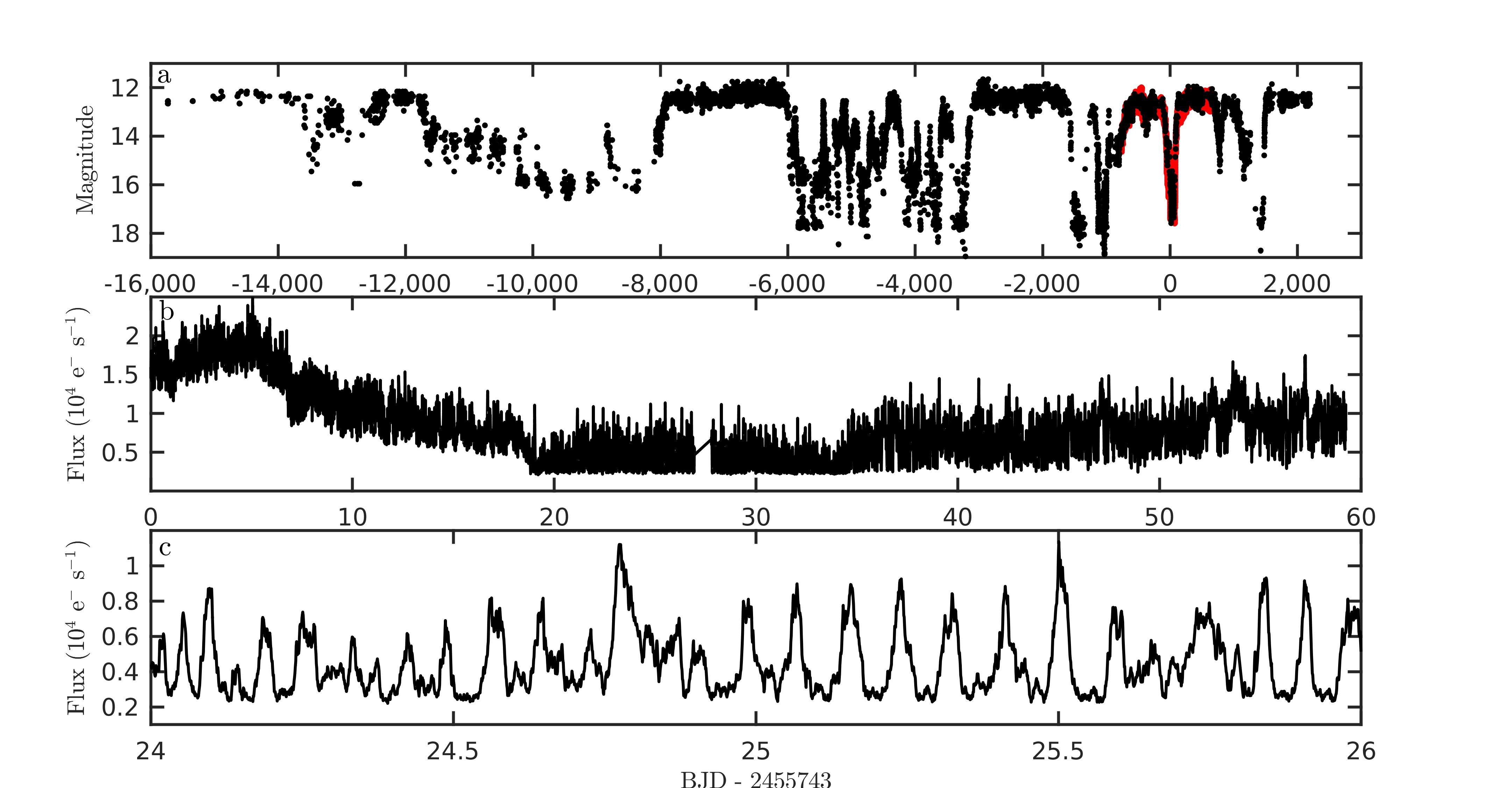}
\caption{
\noindent \textbf{Optical brightness variations in MV Lyrae.}
\textbf{a.} 3.89 year-long \kepler\ lightcurve (58.8 second cadence; red points) overlaid onto the long term V-band and visual lightcurve (black points) obtained through the AAVSO. \textbf{b.} 60-day portion of the \kepler\ lightcurve entering the the deep low state, visible between day 20-35 (BJD - 2455743). \textbf{c.} Further zoom of the \kepler\ lightcurve during the deep low state showing the clear $\approx 30$ minute bursts every $\approx 2$ hours overlaid onto a constant (flat) luminosity level.  
}
\end{center}
\end{figure*}

MV Lyr was continuously monitored during the original \kepler\ mission in short cadence mode (58.8 seconds) for nearly 4 years, displaying both high and low states during this interval (Fig 1a). Although its orbital period has been determined to be 3.19 hours via phase-resolved spectroscopy$^{18}$, the \kepler\ lightcurve does not display any coherent periodicity during the full observation, possibly due to the very low system inclination$^{18,19}$ ($i=10^o \pm 3^o$). Instead the \kepler\ data displays all the usual aperiodic variability patterns that have been associated with mass-transferring accretion disks$^{20-22}$. During an observed low state, MV Lyr displayed quasi-periodic ``bursts'' of $\approx 30$ minutes duration every $\approx 2$ hours, which on occasion can increase the brightness of the system by a factor 6.5 ($\Delta m_V \approx 2$, Fig. 1c). This phenomenon is only observed during the very faintest period of time when the lightcurve reaches a roughly constant minimum brightness level, which we refer to as the deep low state. The bursts are observed only once the brightness of MV Lyr has reached the deep low state, disappearing as soon as the lowest brightness level rises again.

To understand the constant minimum brightness level exhibited in the deep low state, the \kepler\ lightcurve of MV Lyr has been transformed into V-band magnitudes using archival simultaneous observations obtained by the American Association of Variable Star Observers (AAVSO). This allows us to estimate the deep low state brightness level of MV Lyr to be $m_V \approx 17.5$, which is compatible with emission only originating from the white dwarf and secondary star component, with negligible accretion disk contribution. Our estimate is also consistent with previous Far Ultraviolet Spectroscopic Explorer (FUSE) observations of MV Lyr during a previously detected deep low state$^{19,23}$ which did not appear to display any bursting behaviour. The time-averaged magnitude during which MV Lyr reaches the deep low state observed with \kepler\ is $m_V \approx 16.7$. This includes the observed quasi-periodic bursts, and translates to a time-averaged mass accretion rate onto the white dwarf of $\gtrsim 10^{-11} M_{\odot}/\text{yr}$ (see \textit{Methods}).

The combination of the duration, the recurrence time, the large amplitude, and the lack of coherence associated with the quasi-periodic bursts (Fig. 2) exclude an origin of either rotation or pulsation in either the donor star or the white dwarf. One possibility are the simulated Papaloizou-Pringle instabilities generated within the boundary layers of accreting white dwarfs. However, the observed burst recurrence behaviour and the similar burst luminosities cannot be reconciled with current simulations$^{24}$. The most likely mechanism are magnetically-gated accretion bursts, arising from the interaction between the inner edge of the accretion disk and a dynamically important white dwarf magnetic field$^{4,5,25}$. Such bursts can occur when the magnetic field is strong enough to disrupt the disk close to the star, moving the inner edge of the accretion disk outside the ``co-rotation radius'' -- the point where the Keplerian frequency of the disk matches the white dwarf rotation rate. This creates a centrifugal barrier that inhibits accretion onto the white dwarf (Fig. 3). In some cases, the magnetic field is not strong enough to expel most of the accreting gas from the system (as is the case for a ``magnetic propeller''; e.g. AE Aquarii$^{26}$), and as a result, gas in the disk piles up and gradually pushes against the magnetic field (the so-called trapped disk scenario$^5$). Once a critical amount of mass has accumulated, the centrifugal barrier induced by the rotating magnetosphere can be overcome and material accretes onto the white dwarf, releasing a burst of energy through accretion.

\begin{figure}[h]
\begin{center}
\includegraphics[width = 0.5\textwidth]{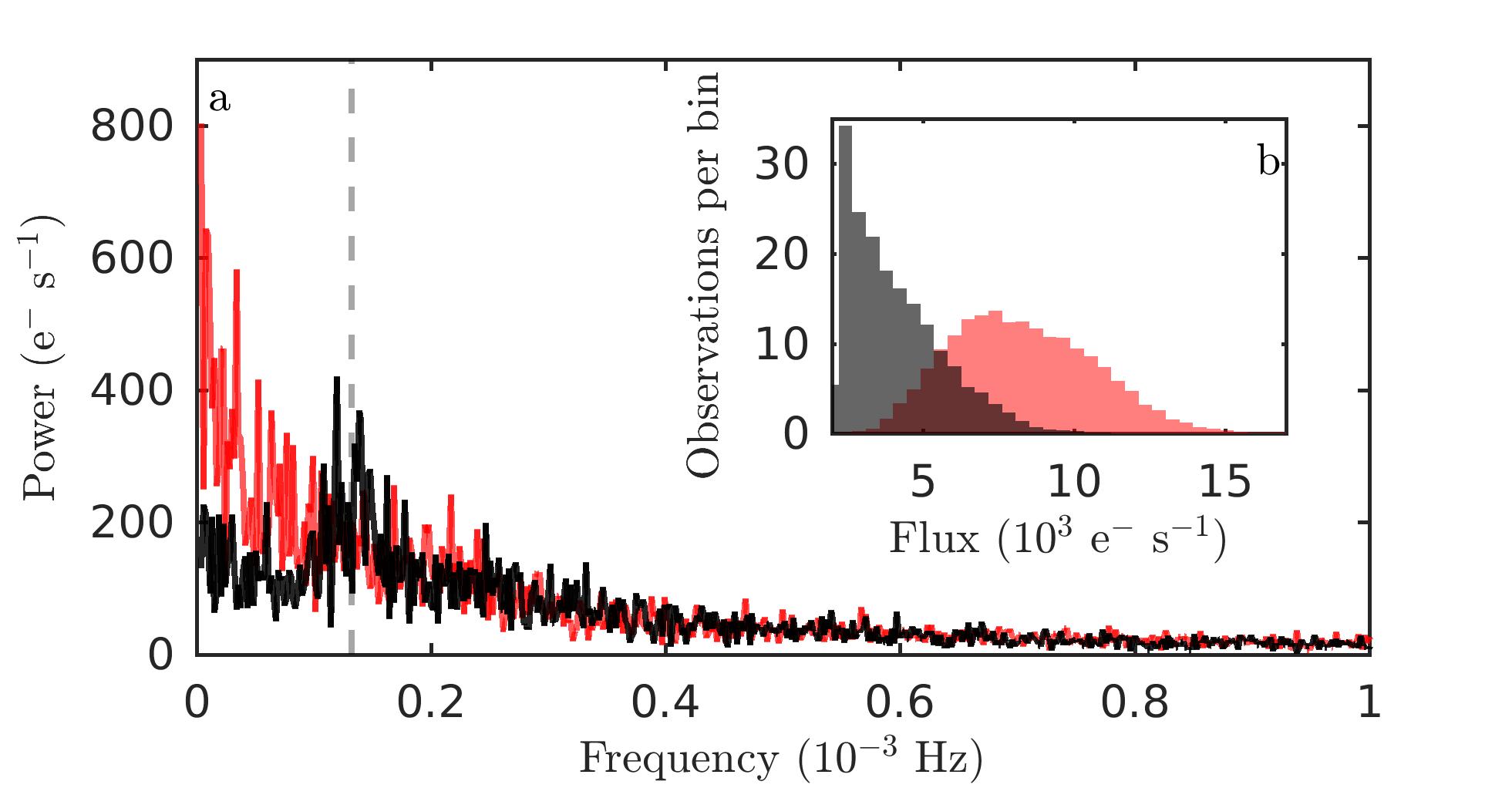}
\caption{
\noindent \textbf{Power spectrum and flux distribution of MV Lyrae in the deep low state and regular low state.}
\textbf{a.} Power spectra during the deep low state lightcurve interval (black line) between 930-944 (BJD-2454833), compared to the interval 955-969 (BJD - 2454833). Due to an observational gap in the \kepler\ lightcurve during the deep low state, we present the average of 2 power spectra, each of equal length of 6.2 days for both the deep low state and the regular low state. The quasi-periodicity is visible in the deep low state as a broad forest of peaks, distinct from the regular low state power spectrum shown red. The dashed gray line marks a 2.1 hour frequency for reference. \textbf{b.} Flux distribution of the corresponding light curve intervals used to compute the power spectra. The deep low state flux distribution (black) is skewed to lower fluxes, abruptly cutting off at $\approx2000$ e$^-$/s. The regular low state flux distribution can be described using a log-normal function, consistent with accretion-induced flickering observed in most accreting sources$^{20-22}$. }
\end{center}
\end{figure}

The critical mass transfer rate required for triggering magnetically-gated accretion burst cycles depends on both the spin period of the white dwarf (and thus the co-rotation radius) and its magnetic field strength (Fig. 4). In the case of MV Lyr, we can constrain the mass transfer rate to be between $10^{-11}\lesssim \dot{M} \lesssim 2\times 10^{-10} M_{\odot}/\text{yr}$. The lower limit arises from the observed time-averaged luminosity in the deep low state. The upper limit arises from the total low state duration of $\approx 300$ days, during which no thermal-viscous outburst was observed$^{27}$ (see \textit{Methods}). Given the inferred mass transfer constraint, we are able to place very conservative constraints on both white dwarf spin period and magnetic field strength by requiring the disk truncation radius to lie between the disk circularisation radius and the white dwarf surface. For a system like MV Lyr ($M_{WD}=0.73\pm 0.1 M_{\odot}$ and $R_{WD}=0.0125 \pm 0.0025 R_{\odot}$), with a 3.19 hour orbital period, the inferred magnetic field of the white dwarf is then constrained to be between 22kG and 1.3MG. The exact value of the field strength depends primarily on the white dwarf rotation period (Fig. 4).

\begin{figure*}[h]
\begin{center}
\includegraphics[width = \textwidth]{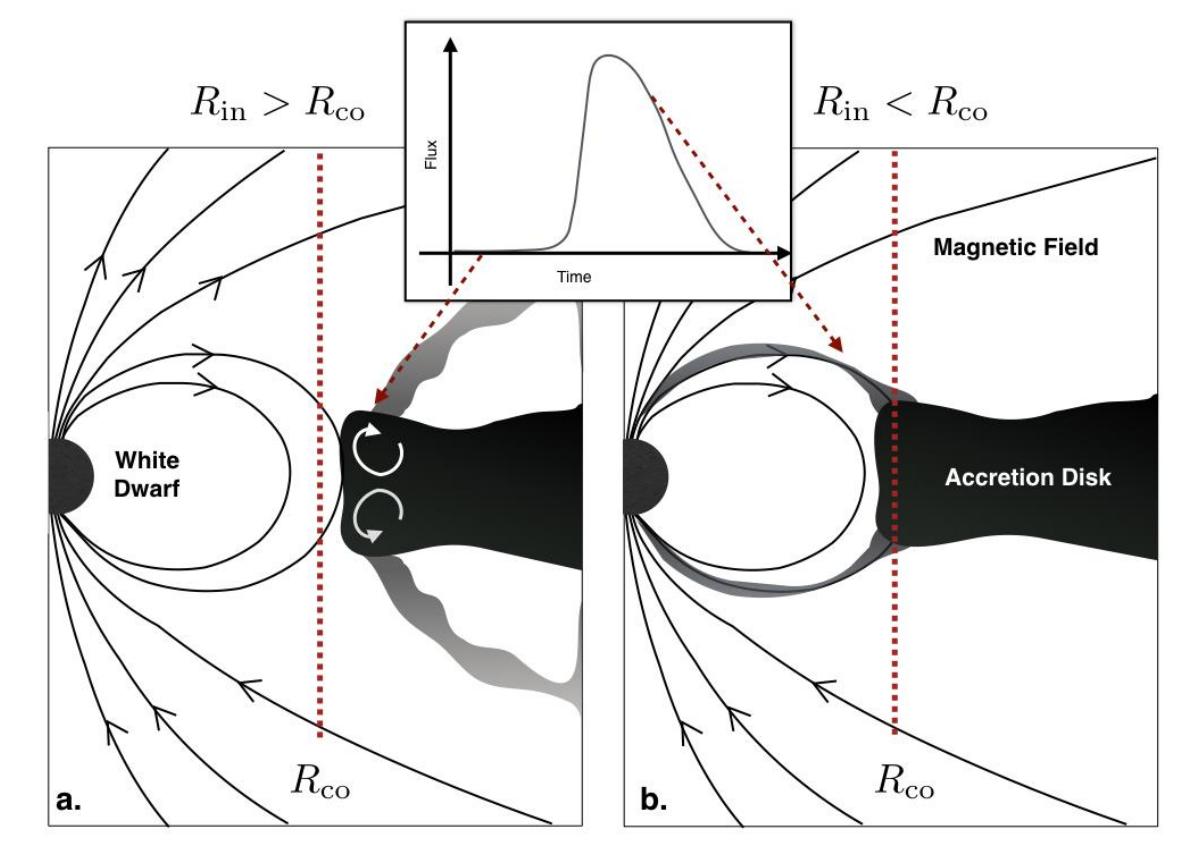}
\caption{
\noindent {\bf Schematic depiction of the accretion flow in MV Lyrae during phases of magnetically gated accretion cycles}
\textbf{a.} During the deep low state accreting gas around MV Lyrae is not able to penetrate the centrifugal barrier created by the interaction between the fast-rotating white dwarf magnetic field and the inner accretion disk. The inner disk is thus truncated just outside the co-rotation radius, preventing the launching of a strong “propeller” outflow. Consequently, material gradually piles up around the truncation radius, exerting more pressure against the magnetic barrier. \textbf{b.} The gas disk eventually pushes inside the co-rotation radius, removing the centrifugal barrier and allowing a burst of accretion onto the white dwarf surface. When the reservoir is depleted, the magnetosphere again pushes outwards and the cycle repeats on timescales comparable to the viscous timescales of the variable disk truncation radius.
}
\end{center}
\end{figure*}

Because MV Lyr is seen nearly face-on, and the white dwarf spin axis is most likely nearly perpendicular to the orbital plane, measurements of the projected white dwarf rotational velocity at the surface are bound to be small, even for a rapid white dwarf spin. High resolution spectra obtained with \textit{FUSE} during a previous low state have been used to infer a projected white dwarf rotational velocity$^{28}$ in the range 150 - 250 km/s. Together with the observed system inclination ($i=10^o \pm 3^o$), this translates to a white dwarf spin period in the range 19-98 seconds, and associate magnetic field strength between 22kG and 130kG. 

The observation of magnetic gating in MV Lyr connects this source to other magnetic accretors, such as young stellar objects and neutron stars, where similar bursts have been seen. For example, EX Lupi, an accreting young star, is the prototype of the ``EXor'' stellar class, which undergo large-amplitude accretion variations that have been attributed to magnetic gating$^{29}$. In EX Lupi, the burst recurrence time of several years corresponds well with viscous timescales in the inner disk region and imply a magnetic field of $\approx 10^3$ Gauss. Observations comparing the inner disk during the accretion burst and post-burst also revealed a depleted inner accretion disk$^{30}$ after the burst. Magnetic gating is also thought to be responsible for very large-amplitude accretion bursts with a $\approx 1$ second recurrence time seen in two different accreting neutron stars with magnetic fields of $\approx 10^8$ Gauss$^{6-8}$. As in MV Lyr, the recurrence time observed in these other accretors is similar to the viscous timescale of the inner disk. The identification of magnetically gated accretion bursts, together with the combination of accretion rate and rotation rate, suggests that the disk is truncated very close to the co-rotation radius at $0.014R_{\odot}<R_{in}<0.043R_{\odot}$ (see \textit{Methods}). By establishing the presence of dynamically important magnetic fields in ``non-magnetic'' white dwarfs, this discovery opens a new route for the study of the strength and evolution of magnetic fields in white dwarfs. Furthermore the new observation of accretion bursts in MV Lyr fill the gap in the magnetic field strength distribution of systems displaying magnetic gating and thus underscores the universality of magnetospheric accretion across an enormous range of stellar parameters$^{6-9,29,30}$.

\begin{figure*}[h]
\begin{center}
\includegraphics[width = \textwidth]{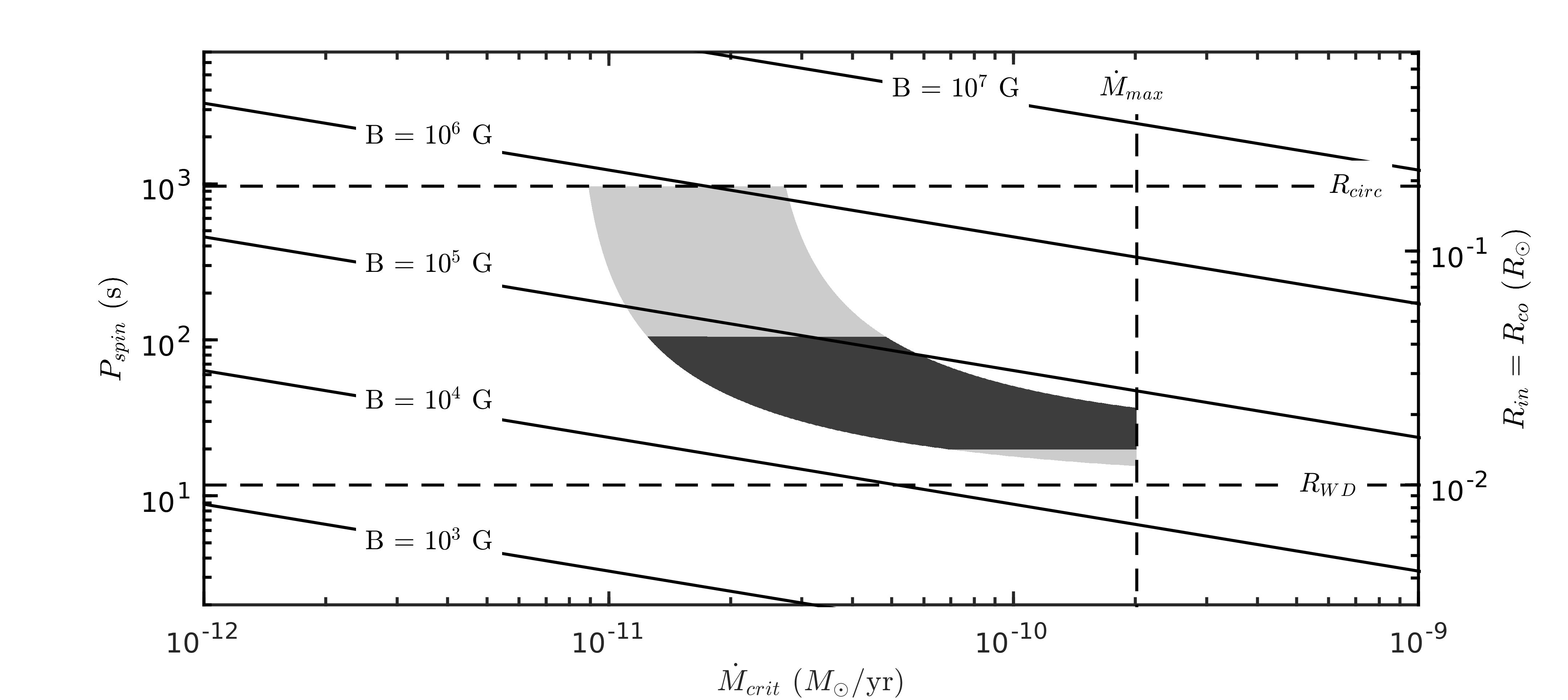}
\caption{
\noindent {\bf Magnetically gated accretion instability plane}
Magnetically gated accretion burst cycles occur when the mass transfer rate through an accretion disk drops to, and is sustained at, a critical value. The critical mass transfer rate ($M_{crit}$) depends on both the magnetic field strength ($B$) and the spin period ($P_s$) of the accretor. This 3-parameter combination determines whether the inner disk ($R_{in}$) is truncated close to the disk co-rotation radius (see \textit{Methods}). The figure shows the MV Lyrae constraints on the critical mass transfer rate and white dwarf spin period (or equivalently inner disk truncation radius), varying the white dwarf magnetic field strength. The horizontal dashed lines mark the strongest constraint given by the outer disk circularisation radius ($R_{circ}$) and the white dwarf radius ($R_{WD}$). The vertical dashed line marks the upper limit on the mass transfer rate in MV Lyr during the $\approx 300$ day low state period observed with \kepler. Our mass transfer rate constraint using the time-averaged flux in the deep low state is shown with the light grey-shaded region. The dark grey-shaded region shows the additional constraint on spin period inferred from the white dwarf rotational velocity$^{28}$ of 150-250 km/s and inclination$^{18,19}$ $i=10^{o}\pm 3^{o}$. The diagonal solid lines mark the critical mass transfer rates at varying white dwarf magnetic field strengths adopting $M_{WD}=0.73 M_{\odot}$ and $R_{WD}=0.0125 R_{\odot}$ for reference.
}
\end{center}
\end{figure*}

\section*{METHODS}

\subsubsection*{Data Sources}
The \kepler\ data for MV Lyr was obtained from The Barbara A. Mikulski Archive for Space Telescopes (MAST) in reduced and calibrated format$^{31,32}$. The \kepler\ telescope/detector combination is sensitive to light across a wide range of wavelengths (400 nm - 900 nm). This maximizes the signal-to-noise and gives robust relative brightnesses over time for the sources, but also makes it difficult to calibrate \kepler\ photometry, either in absolute terms or against other observations obtained in standard (narrower) passbands. In the case of MV Lyr, we are fortunate to have access to the extensive historical data set on this system obtained by amateur observers that is curated by the AAVSO (see Fig. 1a). This data set includes V-band observations spanning nearly 50 years, including the entire period over which \kepler\ observed the source. We have exploited this overlap to establish an approximate transformation of \kepler's count rates for MV Lyr into standard V-band magnitudes. In order to achieve this, we first excluded outliers from the AAVSO data set and removed all observations separated by more than 2 minutes in time from the nearest \kepler\ data point. We then linearly interpolated the \kepler\ light curve onto the timestamps of the remaining AAVSO data and fitted a 6th-order polynomial to the relationship $m_V = f(m_{inst})$, where $m_{inst} = -2.5 log_{10}(\text{count rate}) + 12$ and $f(x) = 13.34 + 1.097x + 0.05851x^2 - 0.08269x^3 + 0.01745x^5 - 0.003327x^6$. By allowing for a higher-order polynomial transformation, we are implicitly correcting for colour terms arising from the difference between the two bandpasses, under the assumption that the colour of the system primarily tracks its luminosity. The root-mean-square scatter about our transformation is 0.07 magnitudes across the entire dynamic range, which spans $12 < m_V< 18$. Due to the increase in noise at the faint end of this range ($m_V > 15.5$), the scatter is slightly higher in this limit (0.14 magnitudes).

\subsubsection*{Power Spectrum}
We visually identified the deep low state of MV Lyr in the \kepler\ lightcurve to fall between 930-944 (BJD-2454833). During this interval a $\approx 20$ hour data gap is present in the \kepler\ data. We thus split the deep low state into two 6.2-day uninterrupted segments (thus avoiding the data gap), and interpolate both on the same 58.8 second time grid. We then perform a Discrete Fourier Transform (DFT) on each independently. For comparison, we performed the same segmentation and analysis to \kepler\ data of MV Lyr just after the deep low state, selecting the segment between 955-969 (BJD - 2454833). Fig. 2a shows the result of averaging the individual deep low state DFTs with the black line, and the comparison DFTs with the red line. We point out that no coherent periodicity is observed in any DFT, with the exception of a known recurrent artefact at $\approx 390$ cycles/day, present in many other \kepler\ targets, and particularly strong during Campaign 10 when MV Lyr entered the deep low state$^{33}$. The fact that no coherent periodicity has been found in nearly 4 years of \kepler\ observation of MV Lyr is also ascertained by numerous other analyses$^{20-22,34,35}$.

\subsubsection*{Luminosity}
During the period in which MV Lyr displays the quasi-periodic bursts we interpret as a signature of magnetically gated accretion, the minimum count rate between the bursts is approximately constant, at a level corresponding to $m_V \approx 17.5$. This stable count rate is therefore most likely due to a combination of light from the white dwarf, the donor star and the accretion flow truncated at radii larger than the co-rotation radius, $R_{in}>R_{cor}$. In order to estimate the time-averaged count rate that is released by the bursts themselves, we therefore first calculate the average count rate across the entire low state displaying the bursts between 930 - 944 (BJD-2454833), and then subtract the stable deep low state minimum count rate. We then use our AAVSO-based calibration to convert the time-averaged burst count rate into an equivalent V-band flux density($f_V \simeq 6 \times 10^{-16}$ erg s$^{-1}$ cm$^{-2}$ \AA$^{-1}$). This, in turn, allows us to estimate the corresponding V-band luminosity via $L_V = 4 \pi d^2 f_V \lambda_{eff,V} \simeq 10^{32}$ erg s$^{-1}$, where $d =505\pm 50$ pc is the distance towards MV Lyr$^{19,23}$, and $\lambda_{eff,V} \simeq 5450$ \AA\ is the effective wavelength of the V-band filter.

\subsubsection*{Accretion Rate}
In the absence of information about the spectral shape of the radiation produced by the bursts, we assume that most of this radiation emerges in the optical region and estimate the burst-related accretion luminosity as $L_{acc} \simeq L_V$, i.e. without making a bolometric correction. We then convert this luminosity into an estimate of the accretion rate onto the white dwarf via

\begin{equation}
L_{acc} = {G \dot{M} M_{WD} \over R_{WD}} \left[ 1 - {1 \over 2} r - \left({\Omega_* \over \Omega_{in}} \right)r + {1 \over 2} \left( {\Omega_* \over \Omega_{in}} \right)^2 r^3 \right] , 
\label{eq:1}
\end{equation}

\noindent where we defined $r = {R_{WD} \over R_{in}}$, $\Omega_*$ is the stellar rotation angular frequency, and $\Omega_{in} = \sqrt{{GM_{WD} \over R_{in}^3}}$ as the inner disk angular frequency. This estimate assumes that $L_{acc}$ represents the gravitational potential energy release associated with material falling from $R_{in}$ to $R_{WD}$ (the surface of the white dwarf). In our estimate we set the inner disk edge to be truncated at the co-rotation radius, such that ${\Omega_* \over \Omega_{in}} = 1$. The expression is derived accounting for the kinetic energy retained by the gas after accreting, as well as the energy lost to spinning up the star$^{36,37}$. Assuming the standard system parameters for MV Lyr$^{19,23,38}$ ($M_{WD} = 0.73\pm 0.1 M_{\odot}$ and $R_{WD} = 0.0125 \pm 0.0025 R_{\odot}$), this yields a lower limit of $\dot{M} \gtrsim 10^{-11} M_{\odot}$ yr$^{-1}$. An upper limit on $\dot{M}$ can also be obtained by noting that during the $\approx 300$ day-long low state observed with \kepler, no thermal-viscous outburst has been observed. The total disk mass required to trigger an outburst$^{27}$ in a system like MV Lyr can be estimated to be $M_{disk} = 1.7 \times 10^{-10} M_{\odot}$. This translates to an upper limit on the mass transfer rate of $\dot{M} \lesssim 2 \times 10^{-10} M_{\odot}$ yr$^{-1}$. The estimates of both $L_{acc}$ and $\dot{M}$ are approximate limits, since we have not made any bolometric correction. If the radiating region is hot, for example, most of the burst energy may be released in the far-ultraviolet or X-ray band rather than in the optical$^{39-42}$, somewhat increasing the $\dot{M}$ estimate.

\subsubsection*{Additional Constraints}
We can set independent limits on the white dwarf spin period, accretion rate and magnetic field strength by setting the magnetospheric radius (where the disk is truncated) equal to the co-rotation radius. When the disk is truncated close to co-rotation, the mass accretion rate can be expressed as$^4$,

\begin{equation}
\dot{M}_{crit} = { \eta \mu^2 P_s \over 8 \pi R_{in}^5 } ,
\label{eq:2}
\end{equation}

\noindent where $\mu = BR_{WD}^3$ is the magnetic moment of the white dwarf with radius $R_{WD}$, $P_s$ is the spin period of the the white dwarf, and $R_{in}$ is the inner disk truncation radius. The time-averaged strength of the toroidal field component, $\eta= {B_{\phi} \over B_z}$, is set to a constant$^{4,5,25}$ of 0.1. Adopting the standard stellar parameters for MV Lyr, we infer $15 <P_s <907$ seconds, $-11.1 < \text{log}_{10} {\dot{M} \over M_{\odot}/\text{yr}} < -9.7$, $4.3<\text{log}_{10} {B\over \text{Gauss}} < 6.1$ and $0.012<R_{in}<0.189 R_{\odot}$, displayed in Fig. 4 as the light grey shaded area. This includes the additional constraint on the outer disk circularisation radius, the inner disk truncation radius at the white dwarf surface, and the allowed maximum mass transfer rate. We can further constrain these parameters through the observed white dwarf projected velocity$^{28}$ of 150-250 km/s and the inferred system inclination of $i=10^o \pm 3^o$. This yields $19<P_s<98$ seconds, $-10.9 <\text{log}_{10} {\dot{M} \over M_{\odot}/\text{yr}} <-9.7$, $4.3<\text{log}_{10} {B\over \text{Gauss}} <5.1$ and $0.014<R_{in}<0.043 R_{\odot}$, displayed in Fig. 4 as the dark grey shaded area.

\subsubsection*{Burst Recurrence Timescales}
In the magnetic gating model, the instability occurs in the inner regions of the accretion disk$^{4,5,25,43}$ and the recurrence time is typically similar to the viscous timescale in this region, (the time it takes matter to travel $R_{in}$; a characteristic evolution timescale):

\begin{equation}
t_{visc} \approx {R_{in}^2 \over \nu } ,
\label{eq:3}
\end{equation}

\noindent where $\nu$ is the viscosity of a typical $\alpha$-disk$^{44}$, and can be estimated as:

\begin{equation}
\nu \approx \alpha (H/R)^2 (GM_{WD}R_{in})^{1/2} .
\label{eq:4}
\end{equation}

\noindent Assuming that the disk is truncated at the co-rotation radius,

\begin{equation}
R_{co} = \left( { GM_{WD} P_{s}^2 \over  4\pi^2 } \right)^{1/3},
\label{eq:5}
\end{equation}

\noindent a white dwarf with $M_{WD} = 0.73 \pm 0.1 M_{\odot}$ and a spin period $19<P_s<98$ seconds has a characteristic viscous accretion time in the range 0.8-4.4 hours. This is consistent with the bursts recurrence timescale observed with \kepler, and possibly with other reports of bursts observed in previous low states$^{45}$. The viscous timescale is calculated assuming a viscous parameter $\alpha \approx 0.1$ and disk aspect ratio $H/R \approx 0.1$, which is plausible if the inner regions of the disc are no longer geometrically thin as is seen at low accretion rates in neutron stars and black holes. Theoretical models of accretion disks around white dwarfs$^{46}$ predict a much lower value for both $H/R$ and $\alpha$. However, several observational results suggest a much higher value of $H/R$ $^{6,20,47}$. The $\alpha(H/R)^2$ parameter is thus somewhat arbitrary, as long as it is not  greater than 1.

\bibliography{scibib}

\bibliographystyle{Science}

\paragraph*{Acknowledgments:}
This paper includes data collected by the Kepler mission. Funding for the Kepler mission is provided by the NASA Science Mission directorate. We acknowledge with thanks the variable star observations from the AAVSO International Database contributed by observers worldwide and used in this research. Some of the data presented in this paper were obtained from the Mikulski Archive for Space Telescopes (MAST). STScI is operated by the Association of Universities for Research in Astronomy, Inc., under NASA contract NAS5-26555. Support for MAST for non-HST data is provided by the NASA Office of Space Science via grant NNX09AF08G and by other grants and contracts. P. J. G. acknowledges support from the Erskine program run by the University of Canterbury.




\end{document}